**[Article Full Title]**

The Role of Free-breathing GRASP MRI in Accurate Phase Matching with 4D-CT for Motion Representation in Liver Cancer Radiotherapy

**[Author Names]**

Junchao Li, M.B. [1, #]; (Email: lee4662084@gmail.com)

Shengqi Chen, B.Eng [3, #]; (Email: shengqichen@bupt.edu.cn)

Guohua Wu, Ph.D [3]; (Email: wuguohua@bupt.edu.cn)

Jianrong Dai, Ph.D [2]; (Email: dai_jianrong@cicams.ac.cn)

Jiayun Chen, Ph.D [2, *]; (Email: grace_chenjy@163.com)

Fei Liu, M.S. [1, *]; (Email: feiliu@tjh.tjmu.edu.cn)

[#] These authors contribute equally as first author.

[*] These authors contributed equally to this work. Jiayun Chen and Fei Liu contributed equally as corresponding authors.

**[Author Institutions]**

[1] Tongji Hospital, Tongji Medical College, Huazhong University of Science & Technology, Wuhan, China

[2] Department of Radiation Oncology, National Cancer Center/National Clinical Research Center for Cancer/Cancer Hospital, Chinese Academy of Medical Sciences and Peking Union Medical College, Beijing, China

[3] School of Electronic Engineering, Beijing University of Posts and Telecommunications, Beijing, China

**[Corresponding Author Name & Email Address]**

Jiayun Chen, Ph.D [2, *]; (Email: grace_chenjy@163.com)

Fei Liu, M.S. [1, *]; (Email: feiliu@tjh.tjmu.edu.cn)

**Abstract**

**Objective**: To determine whether free-breathing golden-angle radial sparse parallel (GRASP) magnetic resonance imaging (MRI) can accurately represent respiratory-induced organ motion in patients with liver malignancies undergoing stereotactic body radiation therapy (SBRT).

**Methods**: A retrospective analysis of 54 patients undergoing liver SBRT was conducted. Four-dimensional computed tomography (4D-CT), the gold standard for motion assessment, was used to characterize the respiratory motion of liver tumors. Image fusion was performed between free-breathing GRASP MRI and each respiratory phase of 4D-CT using an in-house registration program, with fusion quality quantified by maximum cross-correlation coefficient (MCC) analysis. Validation involved two blinded radiation oncologists: one repeated the image fusion using the Eclipse-built-in module, while the other evaluated clinical relevance on a five-point scale.

**Results**: The 50% respiratory phase of 4D-CT achieved the highest fusion quality with free-breathing GRASP MRI, showing no significant differences compared to the 30% ($P = 0.106$), 40% ($P = 0.632$), and 60% ($P = 0.792$) phases. In contrast, fusion quality declined significantly beyond the mid-respiratory window (30%-60% phases), with poor fusion observed at the 0%, 10%, 20%, 80%, and 90% phases ($P < 0.001$). Validation by the radiation oncologists corroborated these findings, with the 50% phase achieving the highest fusion score. Subjective scores remained above 4 for phases 30%-70%, while scores for the remaining phases fell below 4.

**Conclusion**: Free-breathing GRASP MRI cannot independently represent organ motion across all respiratory phases; it accurately characterizes motion only within the mid-respiratory phases (30%-60%), with optimal performance at the 50% phase. When used as a delineation standard in liver SBRT, GRASP MRI should be combined with 4D-CT or other dynamic imaging modalities to ensure comprehensive motion assessment and accurate target volume definition.

## 1. Introduction:

Radiation therapy remains a cornerstone of curative and palliative oncological care, and continual advances in high-precision imaging have markedly enhanced its therapeutic index. Stereotactic body radiation therapy (SBRT) is now established as an effective, non-invasive alternative for patients with primary or secondary liver malignancies who are medically ineligible for surgical resection, image-guided ablation, or trans-arterial chemoembolization [1, 2]. Prospective trials have consistently demonstrated excellent local control, favorable toxicity profiles, and encouraging survival outcomes following liver SBRT [3-9]. Nevertheless, respiratory-induced motion of both tumor and surrounding normal tissue can compromise dosimetric accuracy, risking under-dosage of the gross tumor volume (GTV) and excessive exposure of healthy hepatic parenchyma [10-12]. Consequently, meticulous target-volume delineation and motion-inclusive treatment planning are imperative. Under free-breathing conditions, abdominal respiratory excursion frequently exceeds 1 cm; therefore, motion information derived from state-of-the-art imaging must be incorporated into the internal target volume to prevent geographical miss, ensure robust tumor coverage, and minimise treatment-related toxicity [13, 14].

The sharp dose gradients characteristic of SBRT mandate meticulous delineation of the GTV. Accordingly, magnetic resonance imaging (MRI) is increasingly used, either alone or in combination with CT, for liver tumor delineation due to its superior soft-tissue contrast and ability to distinguish tumors from adjacent hepatic parenchyma [15, 16]. MRI has consequently become the preferred tool for radiotherapy simulation, offering high anatomical fidelity without concomitant ionizing radiation [17]. The broader concept of MRI-guided radiotherapy (MRIgRT) embraces both MR-Linac delivery and MR-only simulation [18-20].

Although breath-hold, respiratory-gated, breath-triggered, and abdominal-compression strategies have been developed to mitigate motion during MRI, their implementation in MRIgRT workflows may be limited by hardware incompatibility [21]. Additionally, free-breathing acquisitions without abdominal belts or compression devices are preferred for radiotherapy simulation, as they maintain the native anatomical configuration of abdominal organs while enhancing patient compliance during prolonged imaging. Among free-breathing strategies, non-Cartesian reconstruction techniques exemplified by the golden-angle radial sparse parallel (GRASP) sequence have revolutionized the acquisition of high-quality images of moving anatomy. GRASP acquires data continuously in radial trajectories, reducing artifacts caused by respiratory motion and improving image quality for patients with regular breathing by retrospectively reconstructing 3D images. The “golden angle” parameter used in GRASP represents the angular increment between each radial line acquired, maximizing the temporal and spatial coverage of the acquired data [22, 23]. Furthermore, compared to conventional T1/T2-weighted MRI combinations for reference volumes, GRASP enables rapid acquisition under free-breathing conditions, thereby improving patient tolerance [23-25].

For fat suppression, mDixon (modified Dixon) techniques, unlike frequency-selective approaches, exploit the chemical shift difference between water and fat. This difference is encoded in the signal phase, enabling robust water-fat separation during post-processing and generating water-only and fat-only images, thereby enhancing tissue contrast and lesion conspicuity [26]. When combined with the mDixon technique, free-breathing GRASP sequences enable simultaneous respiratory motion artifact suppression and water-fat tissue separation imaging, thereby providing a high-quality dynamic imaging foundation for motion management [27, 28]. Consequently, GRASP-mDixon is poised to serve as (i) a delineation tool for pre-treatment planning and (ii) an on-table target-volume surveillance sequence during magnetic-resonance-guided radiotherapy, yet these potential roles remain to be prospectively validated.

Four-dimensional computed tomography (4D-CT) is widely used in radiation therapy planning to characterize respiratory-induced tumor and organ motion. By acquiring CT data over the respiratory cycle and reconstructing phase-correlated datasets, 4D-CT provides phase-resolved three-dimensional information on tumor position throughout breathing [21]. As an established clinical reference for motion assessment in liver radiotherapy [21], 4D-CT enables quantification of patient-specific motion and supports the generation of an internal target volume (ITV) and its subsequent expansion to a planning target volume (PTV), thereby facilitating motion-adaptive liver SBRT plans that minimize the risk of geographical miss and normal-tissue overdose [29, 30]. Previous studies have explored free-breathing GRASP MRI for respiratory motion-resolved abdominal imaging, suggesting its potential value for motion-informed target delineation [31, 32]. However, despite its motion-robust imaging capability, it remains unclear whether free-breathing GRASP MRI represents a respiratory-phase-averaged liver position or corresponds to a specific respiratory phase. In this study, 4D-CT was used as the reference standard for respiratory motion assessment to identify the respiratory phase that best matches GRASP-derived images, thereby clarifying the motion representation characteristics of GRASP and its potential role in hepatic target-volume delineation for MRI-guided liver radiotherapy.

## 2. Methods and Materials

### 2.1 Participants

This retrospective study was approved by the institutional review board (IRB). Patients who underwent SBRT for liver cancer, liver metastasis, or other cancers, with upper abdominal magnetic resonance simulation (MR-Sim) from August 2019 to July 2020, were identified. Patients were included if they underwent upper abdominal MR-Sim with free-breathing GRASP acquisition, utilizing mDIXON T1w axial scans with dual-echo chemical shift imaging (DE-CSI) of the upper abdomen, along with 4D-CT simulation (4D-CT Sim). We identified 56 eligible patients for the study. Two patients were excluded from the study as they had undergone 4D-CT Sim with the abdominal pressure plate. As a result, a total of 54 patients were included in this study, and their demographic characteristics are shown in Table 1.

**2.2 MRI Image Acquisition and Preprocessing**

All patients underwent MR imaging using a dedicated MR-Sim 3.0-Tesla unit (Ingenia CX, Philips Healthcare, Best, the Netherlands) that was equipped with a body coil for radiofrequency transmission and a 32-channel coil for signal reception. RF coil bridges are used to prevent RF coils from deforming the surface anatomy during MR imaging. The imaging was performed according to the clinical protocol. Axial T1w turbo spin echo (T1w-TSE), axial T2-weighted turbo spin echo (T2w-TSE), and GRASP imaging were routinely performed prior to acquiring DWI and other functional imaging sequences. The field-of-view (FOV) for these functional sequences was planned based on the acquired anatomical images to achieve better localization. Additionally, a subset of patients underwent contrast-enhanced magnetic resonance imaging (CE-MRI) and perfusion-weighted imaging (PWI). The key acquisition parameters are shown in Table 2.

The GRASP image acquisition and reconstruction were performed in the MR console using software version R5.7.1. The GRASP reconstruction technique leverages radial k-space sampling based on the golden angle to continuously acquire data during free breathing, combined with parallel imaging to reconstruct high-resolution images from undersampled data [33]. This approach enhances motion robustness and enables retrospective temporal resolution selection, reducing motion artifacts and improving image quality for dynamic MRI acquisitions [23]. GRASP thus provides a reliable imaging solution for moving organs, such as the liver, and has been shown to outperform conventional Cartesian acquisitions in both temporal and spatial resolution [34]. In-plane data (kx; ky) is acquired using a radial profile order, while through-plane data (kz) is acquired using a low-high Cartesian profile order. Each radial profile is acquired within a single turbo field echo (TFE) shot, forming a radial stack-of-stars sampling scheme.

To further enhance image quality and reduce chemical shift effects, chemical shift imaging (CSI) is also integrated into the GRASP image acquisition. CSI corrects the chemical shift artifacts by acquiring four-phase images, including in-phase (IP), out-of-phase (OP), water (W), and fat (F) images, which can assist in tumor identification [35]. The resulting F images can be corrected to reverse the water-fat shift effect, while the W images represent anatomical geometry with no artifacts caused by fat hyper-intensity signal. In particular, for the GRASP image acquisition, mDIXON technology integrated with DE-CSI was applied. This approach utilizes two echo times to differentiate the signal from water and fat molecules in the body. This technique improves image contrast in water images, facilitating better delineation of tumors and surrounding tissues [36].

The corresponding slice levels of the CT and MRI datasets were identified using external markers. Specifically, surface markers were placed during 4D-CT simulation and subsequently aligned to the zero reference plane to reduce setup errors. After that, the axial plane with the largest liver cross-sectional area was selected in the 50% respiratory phase (corresponding to stable end-expiratory state) to minimize the effects of motion artifacts, and its position was then systematically applied to match and extract corresponding slices across all 10 phases of 4D-CT and the MRI images with water phase.

On this basis, the reconstructed images underwent a series of preprocessing steps before being used for image fusion. The specific preprocessing steps included voxel alignment, adaptive cropping, and data normalization. These preprocessing operations were implemented using MATLAB R2021b, in which adaptive cropping was applied to align the image areas of the 4D-CT and MRI data for different patients, considering the differences in size between the 2D CT images and the reconstructed MRI images, as well as individual variability. Meanwhile, in order to eliminate biases caused by variations in scanning equipment, scanning conditions, or patient differences and to enhance image contrast, normalization was applied to the input, as shown in Equation (1):

$$\begin{aligned} MRI_{i,j}^{norm} &= \frac{MRI_{i,j} - mean(MRI)}{mean(MRI)} \\ CT_{p,i,j}^{norm} &= \frac{CT_{p,i,j} - mean\left(\sum_p CT_p\right)}{mean\left(\sum_p CT_p\right)} \end{aligned} \tag{1}$$

Here, $MRI$ and $CT$ represent the original images, while $MRI^{norm}$ and $CT^{norm}$ represent the normalized images. $i$ and $j$ denote the row and column indices of a single pixel, respectively, and $p$ represents different respiratory phases (from 0% to 90%). $\sum(\cdot)$ refers to the summation operation. From the perspective of the image fusion objective, these necessary preprocessing steps ensure the consistency of the input images, enhance image quality, and the improved contrast helps to further highlight key structures and lesion areas. The complete workflow of the proposed method is shown in Figure 1.

### 2.3 Image Registration and Maximum Cross-correlation Coefficient (MCC) Analysis

We implemented MRI-CT image fusion using an in-house MATLAB R2021b-based image registration program. As shown in Figure 1, preprocessed images were used as input. For different respiratory phases, the cross-correlation matrices between CT and MRI images were computed. The maximum cross-correlation coefficient (MCC) within each matrix was determined, while the translational offset between the corresponding optimal matching coordinates and the image centroids was also calculated. Such an offset was then applied to align CT images for optimal alignment. The adopted method, known as “template matching”, has demonstrated effectiveness in image registration tasks across imaging modalities such as orthophoto-map [37] and radar [38], especially validated in medical imaging applications [39]. The MCC value quantitatively characterizes optimal matching performance and similarity [40, 41], establishing its validity as the primary criterion for evaluating fusion quality. Following normalization to eliminate scale differences, data were categorized by respiratory phase. Subsequent statistical analysis ultimately identified the optimal respiratory motion phase of 4D-CT images for fusion with free-breathing MR images.

### 2.4 Validation

In order to ensure the accuracy of the MCC analysis results and further support our findings, a radiotherapy physician with skilled clinical practice performed automatic image registration using the Eclipse-built-in registration module (ROI uses the range of the patient including the entire liver). All image fusion results were independently reviewed by another radiotherapy physician in random order. The reviewers were blinded to information about the 4D-CT phase.

The quality of the fused images was evaluated using a five-point scale method [42], based on the following criteria for overall fusion quality: (1) Unacceptable—MRI image unsuitable for target delineation; (2) Poor—MRI image not recommended for target delineation; (3) Moderate—MRI image acceptable for target delineation with limited reliability; (4) Good—MRI image suitable for target delineation with high reliability; (5) Excellent—MRI image ideal for target delineation with complete reliability.

### 2.5 Statistical Analysis

The Shapiro-Wilk test was employed to evaluate data normality. In cases with normal distributions, the overall statistical characteristics were expressed as mean ± standard deviation, whereas non-normally distributed data were represented by median (interquartile range). The Mann-Whitney U test was utilized to examine the null hypothesis of no significant difference in image fusion quality between different respiratory phases ($P$-value < 0.05 was considered statistically significant), thereby identifying the optimal respiratory phase for CT-MRI image fusion. All statistical analyses were performed using SPSS version 27.0.1.0 (IBM, Armonk, NY, USA).

## 3. Results

### 3.1 Imaging Results

In this study, a total of 54 patients' upper abdominal free-breathing GRASP with mDIXON T1w axial scans and 4D-CT Sim images were collected. Figure 2 illustrates typical free-breathing GRASP with mDixon T1-weighted axial scans, including four reconstructed image sets: water, in-phase, out-of-phase, and fat. The water-only image (Figure 2(a)) demonstrates excellent fat suppression and clearly delineates the liver parenchyma, vessels, and tumor margins, providing a strong foundation for subsequent image registration and target delineation. The in-phase image (Figure 2(b)) presents combined water and fat signals with uniform signal intensity, allowing for comprehensive anatomical evaluation. The out-of-phase image (Figure 2(c)) exhibits partial signal cancellation between water and fat components, which facilitates the detection of hepatic fat infiltration or mixed tissue composition. The fat-only image (Figure 2(d)) highlights subcutaneous and perihepatic fat structures, assisting in the differentiation between hepatic parenchyma and surrounding adipose tissue.

Overall, the GRASP sequence integrated with the mDixon technique under free-breathing conditions achieved a high signal-to-noise ratio and excellent water-fat separation. This combination effectively minimized respiratory motion artifacts while maintaining consistent image quality across patients, providing reliable soft-tissue contrast for accurate 4D-CT fusion and respiratory motion analysis.

### 3.2 Image Fusion Results

Figure 3 shows the normalized MCC results of template matching for image fusion performed on free-breathing GRASP mDIXON T1w axial scans and 4D-CT images of 54 patients across 10 respiratory phases. The MRI and CT images showed the best fusion effect at the 50% respiratory phase, followed by the 60%, 40%, 30%, and 70% phases. Statistical analysis showed that both the normalized MCC and subjective score were not normally distributed. The statistical normalized MCC values were 10.10% (0.11%) for the 50% phase, 10.09% (0.11%) for the 60% phase, 10.08% (0.10%) for the 40% phase, 10.07% (0.10%) for the 30% phase, and 10.05% (0.08%) for the 70% phase.

Compared to the 50% respiratory phase, no statistically significant decline in image fusion quality was observed at 60% ($P$=0.792), 40% ($P$=0.632), or 30% phases ($P$=0.106). The strong concordance in 30%-60% phases supports the validity of free-breathing GRASP MRI for motion representation within this range. In contrast, significant differences were identified at the 0.05 confidence level when compared with both the 70% phase ($P$=0.013) and other respiratory phases ($P$<0.001), indicating diminishing registration accuracy beyond the optimal mid-respiratory window and highlighting limitations in representing some motion patterns.

Figure 4 shows the subjective scoring results of image fusion performed by a radiotherapy physician using the Eclipse software, demonstrating a similar trend to the normalized MCC results. For the overall fusion quality parameters, the 50% phase of 4D-CT achieved the highest scores, followed by the 60%, 40%, 70%, and 30% phases. The median scores for overall fusion quality were 5 (0) for the 50% phase, 5 (1) for the 60% phase, 5 (1) for the 40% phase, 4 (0) for the 70% phase, and 4 (0) for the 30% phase, 3.5 (1) for the 20% phase, 3 (1) for the 80% phase, 2(1) for the 10%, 90% and 00% phases. Statistical analysis showed that all phases yielded a significant difference in fusion quality from the 50% phase ($P$<0.001).

## 4. Discussion

Respiratory motion can lead to insufficient dose delivery to target volumes and excessive dose to healthy tissues, underscoring the critical role of effective motion management in ensuring target coverage and minimizing radiation-induced damage to healthy organs during SBRT [43]. MRI has emerged as the preferred modality for radiotherapy simulation due to its superior soft-tissue contrast and absence of ionizing radiation, representing the trend of precision radiotherapy [17]. Motion compensation techniques for MRI scanning require MR-compatible hardware and are not universally applicable to all patients. Rapid MRI acquisition is particularly essential in the MR-guided radiotherapy (MRgRT) workflow or on dedicated MRI-simulation platforms, as it reduces the time patients must remain in a fixed position, thereby improving comfort and compliance. Additionally, shortened scan times decrease overall treatment duration in MRgRT, enhancing workflow efficiency. In this context, free-breathing MRI integrated with the GRASP method—which employs continuous radial trajectory data acquisition to suppress respiratory motion artifacts—demonstrates inherent robustness to motion variations. This approach holds promise as a target delineation tool in MR-guided radiotherapy, offering a reliable foundation for motion-compensated radiotherapy planning.

In liver SBRT, the fusion of GRASP-generated T1w MRI sequence with 4D-CT images has been demonstrated to accurately characterize tumor motion during radiotherapy and optimize treatment planning, thereby reducing toxicity doses and enhancing therapeutic efficacy [21]. In this study, 54 patients undergoing liver SBRT were enrolled. Each patient's upper abdominal free-breathing GRASP with mDIXON T1w axial scans and 4D-CT Sim images were fully acquired. The reconstructed imaging results demonstrated that the mDIXON technique integrated with DE-CSI significantly improved liver-to-lesion contrast through precise water-fat signal separation. Compared to conventional fat-suppression methods, the W images generated by mDIXON effectively eliminated chemical shift artifacts, enhancing the delineation clarity between hepatic parenchyma and tumor boundaries. Simultaneously, the F images clearly visualized perihepatic fat interfaces via reversed water-fat shift correction, providing reliable morphological information for target delineation. These findings align with the documented advantages of mDIXON technology in existing literature [36]. Furthermore, combined with GRASP's radial sampling and non-uniform Fourier reconstruction, the mDIXON technique exhibited strong robustness against respiratory motion artifacts under free-breathing conditions. Whether GRASP can accurately represent respiratory-induced organ motion in patients with liver malignancies undergoing stereotactic body radiation therapy (SBRT) constitutes the central focus of this study.

Building on this, to evaluate whether GRASP fully represents organ motion, this study employed the fusion of free-breathing GRASP MRI with individual phases of 4D-CT to investigate which respiratory phases are characterized. A template matching method based on maximum cross-correlation (MCC) analysis was implemented to align GRASP MRI with CT images across different respiratory phases, with subsequent independent manual validation. Data from 54 patients undergoing liver SBRT were enrolled. Experimental results demonstrated that the 50% phase CT images corresponding to end-respiratory achieved optimal fusion quality, with a normalized MCC of 10.10% (0.11%). Significantly, the 30%, 40%, and 60% phases exhibited comparable matching performance (P > 0.05), indicating that GRASP MRI most accurately represents the anatomical characteristics of 4D-CT images during mid-respiratory phases. Meanwhile, the independent clinical validation reinforced these findings. For overall fusion quality assessment, the 50% phase achieved the highest subjective score of 5 (0), demonstrating complete reliability for target delineation. Significant differences were identified when compared to all other phases (P < 0.001). Notably, the 30%, 40%, 60%, and 70% phases maintained scores above 4 (high reliability), confirming the robustness of GRASP MRI in representing tumor motion across mid-respiratory phases.

Compared to the end-expiratory phase, the normalized MCC values for the 0%, 10%, 20%, 80%, and 90% phases exhibited a significant decline (P < 0.001), indicating limitations in the motion representation capability of GRASP MRI. Furthermore, the 0%, 10%, and 90% phases received subjective scores of 2 (1), reflecting poor fusion quality that is clinically unsuitable for target delineation. These findings suggest that the GRASP MRI sequence may not accurately capture motion across all respiratory phases, particularly during the end-inspiratory phase. This may be because the free-breathing GRASP acquisition primarily reflects the average target position over the respiratory cycle rather than the extreme inspiratory or expiratory states. Therefore, its motion representation is expected to be better confined to the mid-respiratory phases (30%-60%). Accordingly, although GRASP MRI demonstrates superior anatomical characterization capability and motion robustness within this range, its limitations in capturing comprehensive respiratory dynamics necessitate integration with pre-treatment 4D-CT. This combination provides a more complete reference for target volume definition with motion-inclusive margins and organs at risk delineation. Our results suggest that fusing MRI sequences with 4D-CT for precise motion detection should be a recommended approach in clinical practice.

In this study, the classical template matching method was employed to achieve image fusion between GRASP MRI and 4D-CT, aiming to identify the most compatible respiratory phase. The peak of the cross-correlation matrix, i.e., the MCC value, quantified the similarity between the registered images [40, 41] and was thus utilized for the quantitative assessment of image fusion quality. The inherent modality-specific differences between MRI and CT necessitate preprocessing steps, including voxel alignment for spatial coordinate correction, adaptive cropping for anatomical region matching, and normalization to mitigate intensity variations and optimize structural contrast. To ensure the robustness of the conclusions, we further adopted the manual validation by expert radiation oncologists as the clinical standard. The results demonstrated that the effectiveness of MCC analysis was supported by the validation result, as it exhibited a phase-dependent variation trend consistent with the subjective scores. This consistency confirmed that the MCC analysis method aligns with clinical practical needs and can serve as a reliable tool for evaluating the quality of MRI-CT image fusion.

The limitations of this study include potential selection bias inherent to the retrospective design, which may compromise the generalizability of the conclusions. Additionally, the single-center cohort comprising 54 patients may not fully represent the statistical heterogeneity of broader populations and could have been influenced by institutional-specific practices. Future work should focus on validating the generalizability of our findings through larger-scale, multi-center prospective trials. Concurrently, the development of more objective quantitative evaluation tools is required to minimize human subjectivity, thereby providing technical support for the clinical application of GRASP MRI.

## 5. Conclusion

Free-breathing GRASP MRI cannot independently represent liver motion across the entire respiratory cycle; it accurately characterizes motion only within the mid-respiratory phases (30%-60%) of 4D-CT, with optimal agreement at the 50% phase. Beyond this window, fusion quality declines significantly, indicating limited motion representation capability at end-inspiratory and extreme respiratory phases. Therefore, when GRASP MRI is used as a delineation reference in liver SBRT, it should be integrated with 4D-CT or other dynamic imaging modalities to ensure comprehensive respiratory motion assessment and accurate target volume definition.

## Figure Captions

Figure 1. Workflow of the proposed image fusion and validation framework.

Figure 2. Imaging results of GRASP with mDIXON T1w axial scans: water-phase(a), in-phase(b), out-of-phase(c), fat-phase(d).

Figure 3. Normalized MCCs of 54 patients across 10 respiratory phases.

Figure 4. Subjective scores of 54 patients across 10 respiratory phases.

Step 1: Image Acquisition and Preprocessing

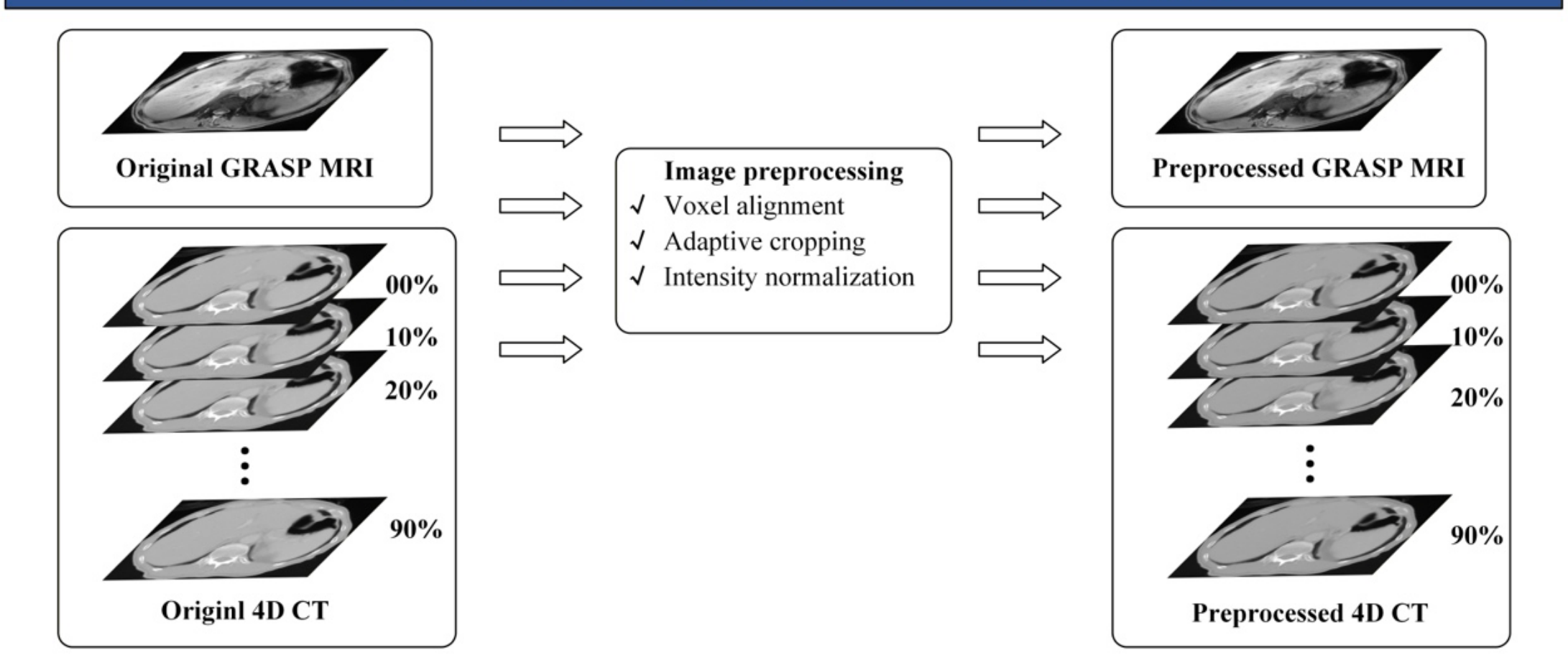

Step 2: Image Registration and MCC Analysis

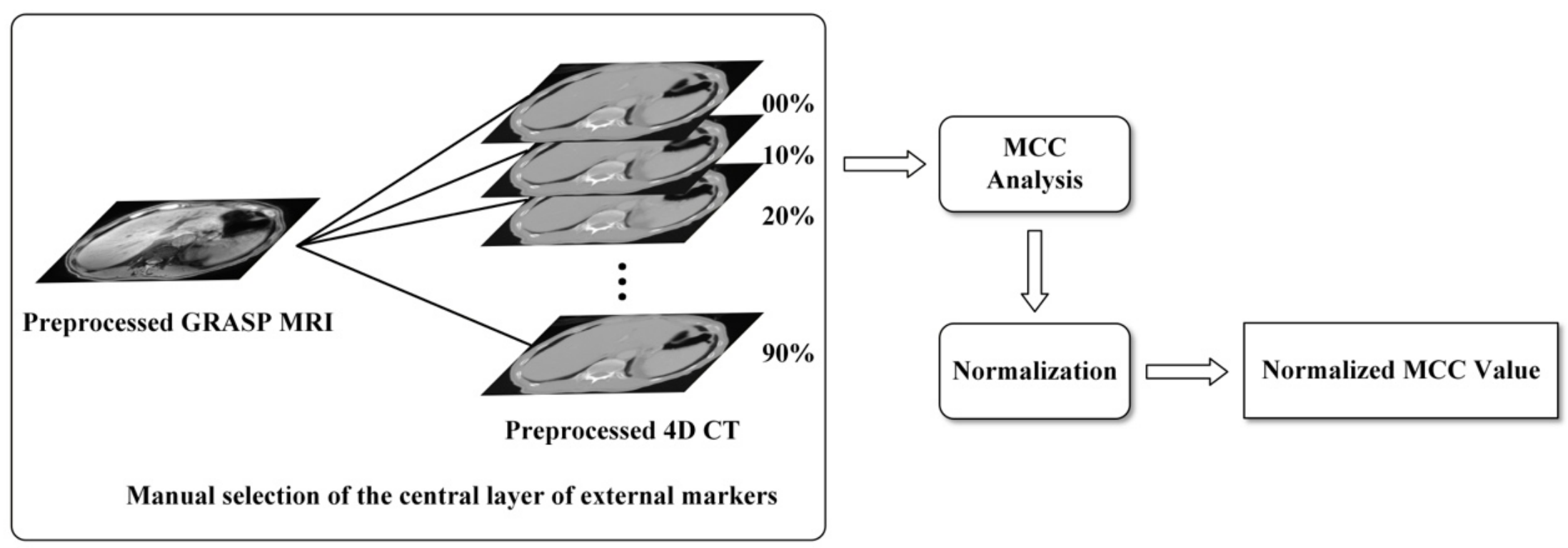

Step 3: Validation

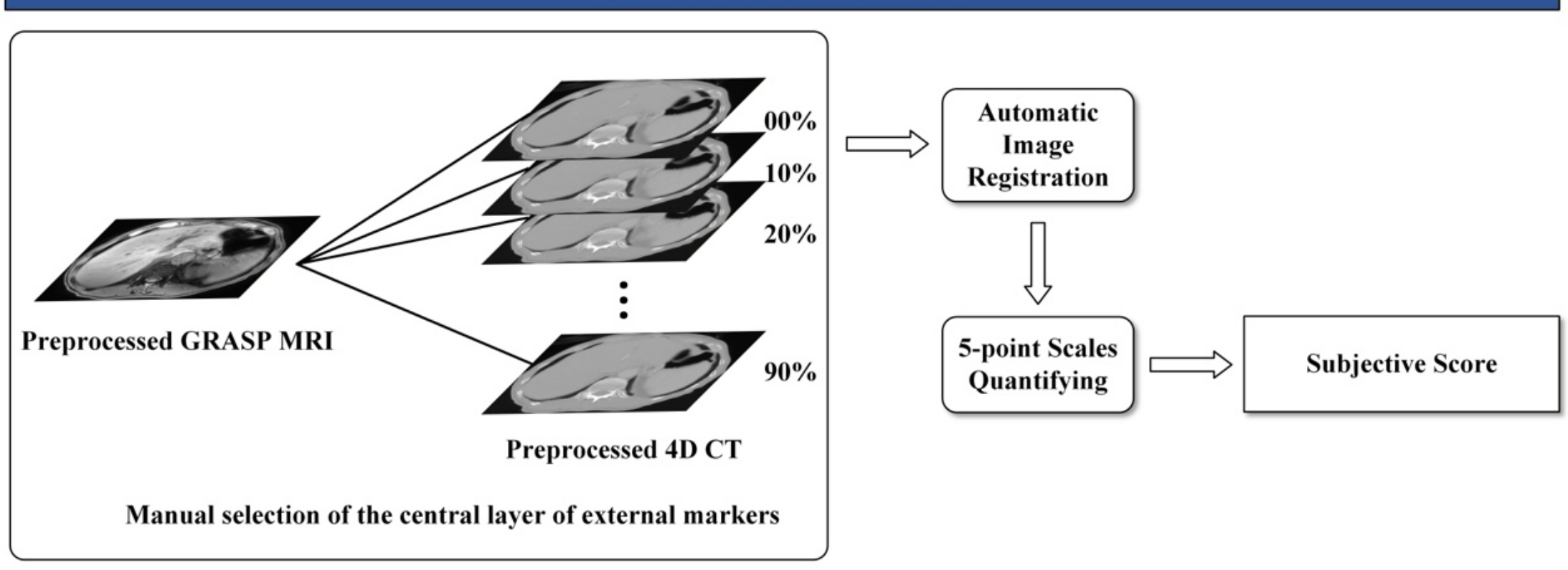

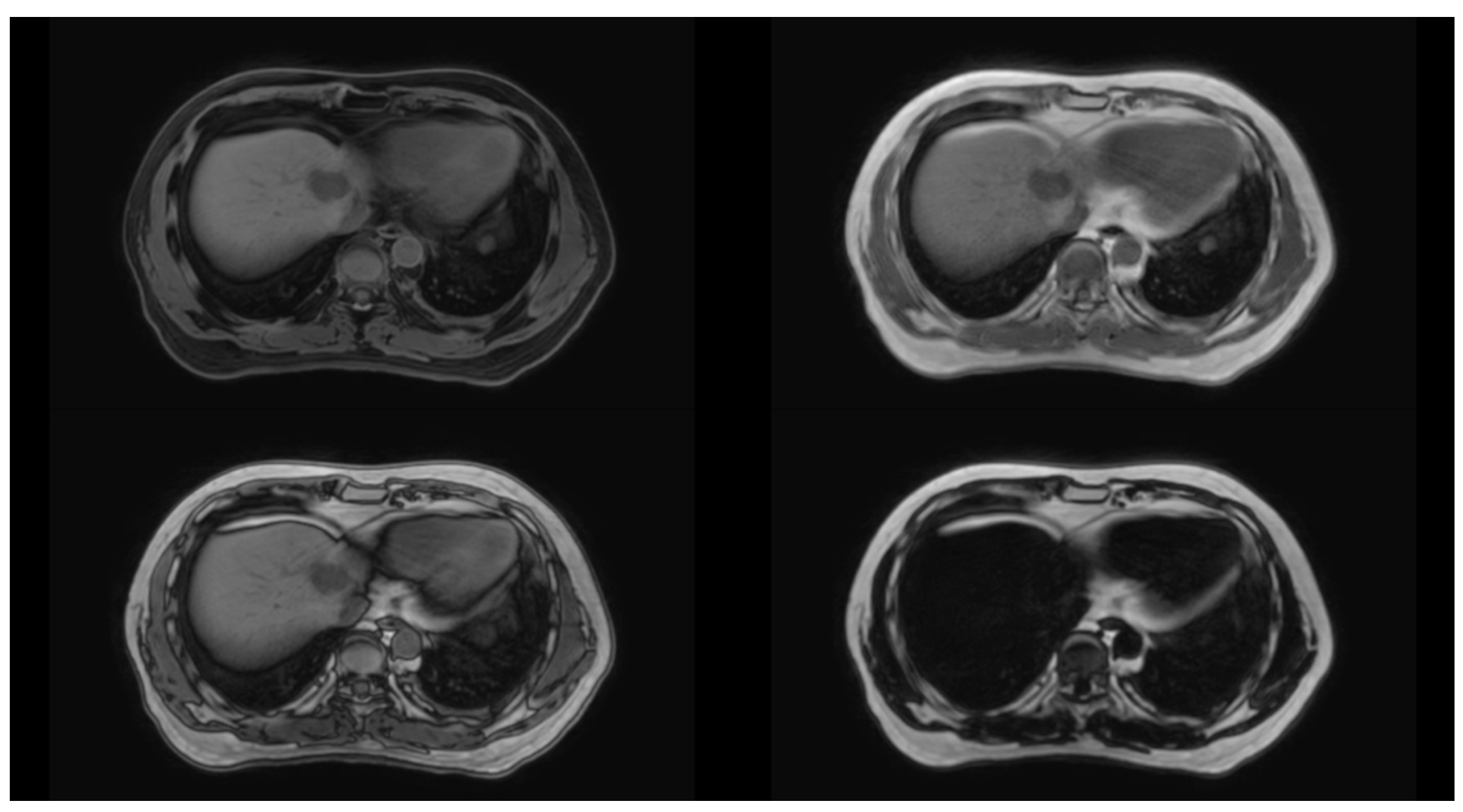

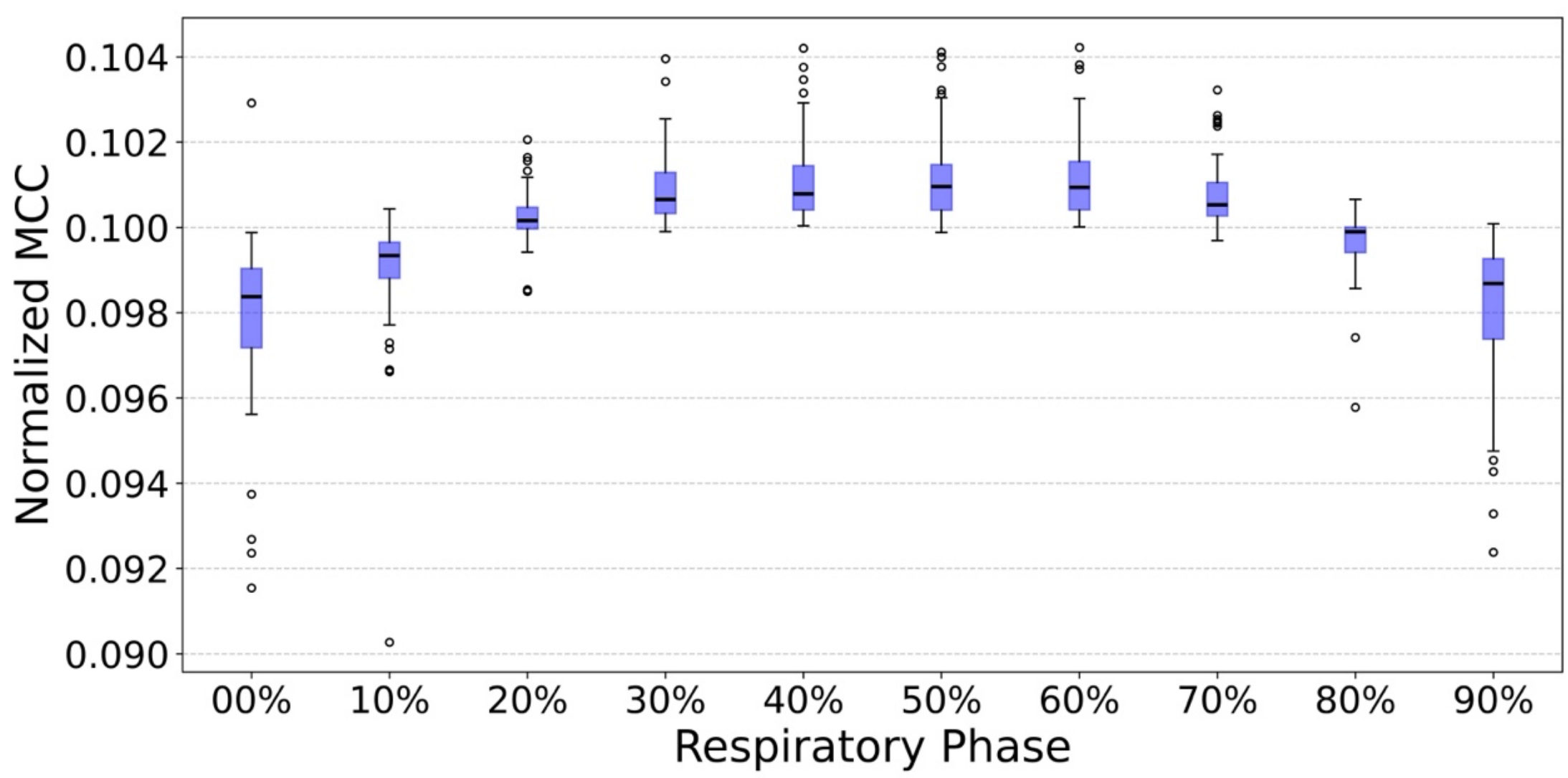

0.104
0.102
0.100
0.098
0.096
0.094
0.092
0.090
Normalized MCC
00%
10%
20%
30%
40%
50%
60%
70%
80%
90%
Respiratory Phase

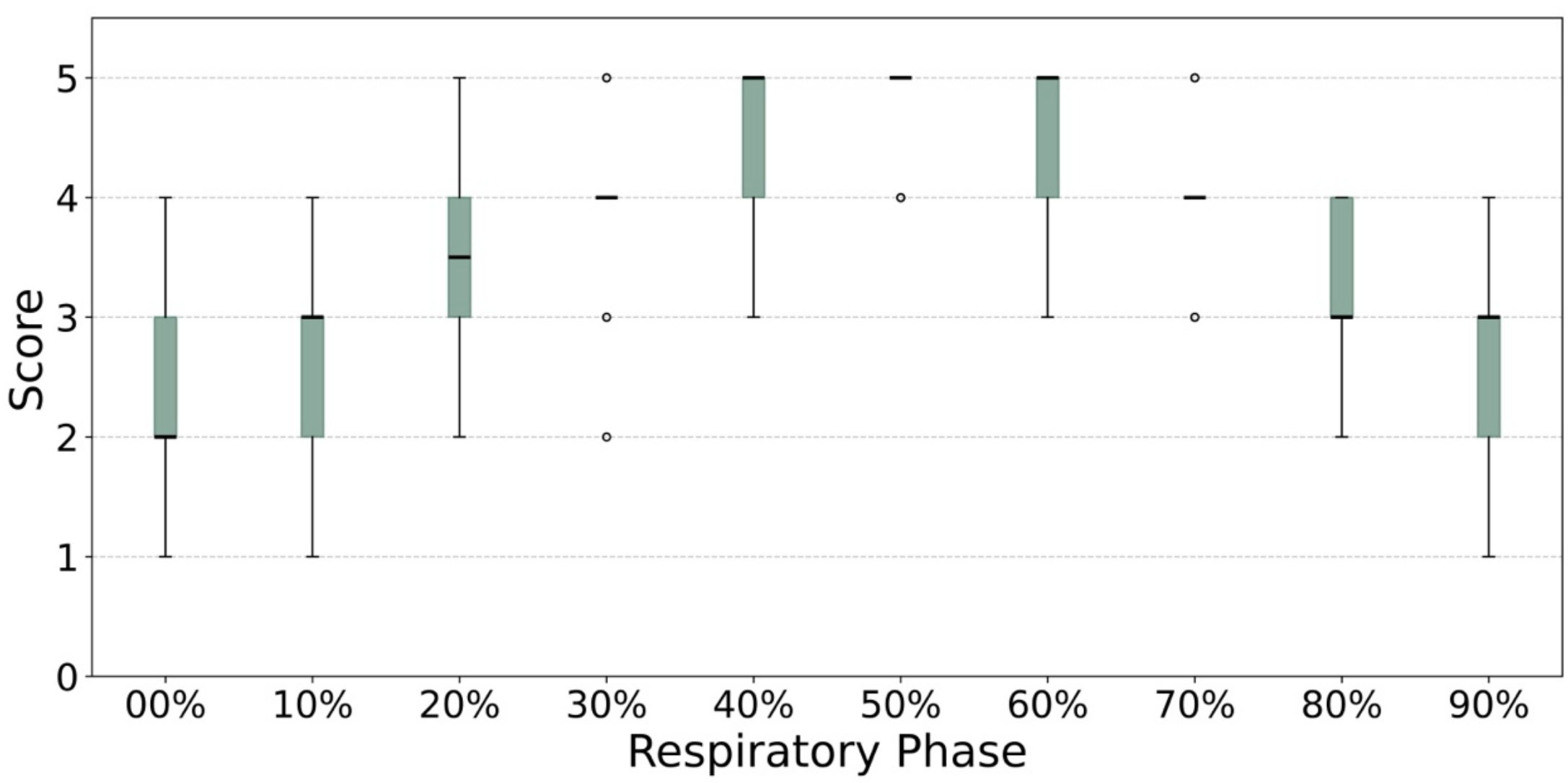
Score
5
4
3
2
1
0
00%
10%
20%
30%
40%
50%
60%
70%
80%
90%
Respiratory Phase